\documentclass[aps,prl,twocolumn,groupedaddress,showpacs]{revtex4}

\usepackage{graphicx}
\usepackage{dcolumn}
\usepackage{bm}
\usepackage{float}

\def\bc{\begin{center}}
\def\ec{\end{center}}

\def\beq{\begin{equation}}
\def\eeq{\end{equation}}

\begin{document}


\title{Random gap model for graphene and graphene bilayers}

\author{K. Ziegler}%
\email{Klaus.Ziegler@Physik.Uni-Augsburg.de}
\affiliation{
Institut f\"ur Physik, Universit\"at Augsburg, D-86135 Augsburg, Germany
}

\date{\today}

\begin{abstract}
The effect of a randomly fluctuating gap, created by a random staggered potential,
is studied in a monolayer and a bilayer of graphene. The density of states, the one-particle
scattering rate and transport properties (diffusion coefficient and conductivity)
are calculated at the neutrality point. All these quantities vanish at a critical value of the average
staggered potential, signaling a continuous transition to an insulating behavior.
The calculations are based on the self-consistent Born approximation for the 
one-particle scattering rate and a massless mode of the two-particle Green's 
function which is created by spontaneous symmetry breaking. 
Transport quantities are directly linked to the one-particle scattering rate. 
Moreover, the effect of disorder is very weak in the case of a monolayer but much stronger 
in bilayer graphene.
\end{abstract}

\pacs{81.05.Uw,71.55.Ak,72.10.Bg,73.20.Jc}
\maketitle


Graphene, a sheet of carbon atoms, or bilayer graphene are semimetals with good 
conducting properties \cite{novoselov05,zhang05,geim07}. In particular, the minimal conductivity
at the neutrality point (NP) is very robust and almost unaffected by disorder or thermal fluctuations
\cite{geim07,tan07,chen08,morozov08}.
Recent experiments with hydrogenated graphene \cite{elias08} and biased bilayer graphene
\cite{ohta06,oostinga08,gorbachev08} have revealed that a staggered potential (SP) can be
created in graphene and bilayer graphene which breaks the sublattice symmetry.
This opens a gap at the Fermi energy, leading to an insulating behavior. 
With this opportunity one enters a new field, where one can switch 
between a conducting and an insulating regime of a two-dimensional material, 
either by a chemical process (e.g. oxidation or hydrogenation) or by applying an 
external electric field \cite{castro08}.

It is clear that the opening of a uniform gap destroys the metallic state immediately.
This means that the (minimal) conductivity at the NP drops from a finite value
of order $e^2/h$ directly to zero. In a realistic system, however, the gap may not
be uniform. This means that locally gaps open, whereas in other regions of the sample there is
no gap. The situation can be compared with a classical random network of broken and unbroken bonds. 
The conductivity of such a network is nonzero as long as there is a percolating cluster of unbroken
bonds. In such a system the transition from conducting to insulting behavior is presumably
a second order percolation transition \cite{cheianov07}.

Disorder in graphene has been the subject of a number of recent numerical studies 
\cite{xiong07,zhang08}. The results can be summarized by the statement that chiral-symmetry 
preserving disorder provides delocalized states whereas a chiral-symmetry breaking potential 
disorder leads to Anderson localization, even at the NP.

Conductivity and other transport properties in graphene can be evaluated by solving 
the Bethe-Salpeter equation for the average two-particle Green's function (Cooperon) 
\cite{suzuura02,peres06,khveshchenko06,mccann06,yan08}. Unfortunately, the Bethe-Salpeter equation 
is usually a complex matrix equation which is difficult to handle. Therefore, a different
approach will be employed here that eliminates a part of the complexity by
focusing on continuous symmetries and spontaneous symmetry breaking. This allows us to identify a
(massless) diffusion mode in the system with a randomly fluctuating gap.
Consequently, diffusion can only stop when the spontaneous symmetry breaking
vanishes. It will be discussed in this paper that this can happen if the average
SP approaches a critical value. Moreover, there is no drop of the conductivity
but a continuous decay to zero, depending on the fluctuations of the SP.

{\it model:}
Quasiparticles in monolayer graphene (MLG) or bilayer graphene (BLG) 
are described in tight-binding approximation by a nearest-neighbor hopping Hamiltonian 
\beq
{\bf H}=-t\sum_{<r,r'>}c^\dagger_r c_{r'}+\sum_r m_r c^\dagger_r c_r +h.c.
\ ,
\label{ham00}
\eeq
where the underlying structure is either a honeycomb lattice (MLG) or 
two honeycomb lattices with Bernal stacking (BLG).

The sublattice symmetry of the honeycomb lattice is broken by a staggered potential (SP)
$m_r$ which is positive (negative) on sublattice A (B) \cite{mccann06b,koshino06}. Such a potential can be
the result of chemical absorption of other atoms (e.g. oxygen or hydrogen \cite{elias08})
or of an external gate voltage applied to the two layers of BLG \cite{ohta06}. 
Neither in MLG nor in BLG the potential $m_r$ and, therefore, the gap is uniform, 
because of fluctuations in the coverage of the MLG by additional non-carbon atoms or 
by the fact that the graphene sheets are not planar \cite{morozov06,meyer07,castroneto07b}. 
Deviations from the planar structure in the form of ripples cause fluctuations in the
distance of the two sheets in BLG which results in an inhomogeneous potential
$m_r$ along each sheet. It is assumed that the gate voltage is adjusted at 
the NP such that in average the SP is exactly antisymmetric:
$\langle m_A\rangle=-\langle m_B\rangle$.

At first glance, the Hamiltonian
in Eq. (\ref{ham00}) is a standard hopping Hamiltonian with random potential $m$, 
frequently used to study the generic case of Anderson localization \cite{anderson58}. The dispersion,
however, is special in the case of graphene due to the honeycomb lattice: at low energies it 
consists of two valleys $K$ and $K'$ \cite{castroneto07b,mccann06}. 
It is assumed that weak disorder scatters only at small momentum such that 
intervalley scattering is not relevant.
Then each valley contributes separately to transport, and the contribution of 
the two valleys to the conductivity $\sigma$ is additive:
$
\sigma=\sigma_K+\sigma_{K'}
$.
This allows us to consider for the low-energy properties a Dirac-type Hamiltonian 
for each valley separately
\beq
H=h_1\sigma_1+h_2\sigma_2+m\sigma_3 
\label{ham01}
\eeq
with Pauli matrices $\sigma_j$ and with $h_j$
\beq
h_j=i\nabla_j \ \ (MLG), \ \ h_1=\nabla_1^2-\nabla_2^2, \ h_2=2\nabla_1\nabla_2 \ \ (BLG) \ .
\label{elements}
\eeq
$\nabla_j$ is the lattice difference operator in $j$ ($=1,2$) direction.
Within this approximation the SP $m_r$ is a random variable with mean
value $\langle m_r\rangle_m ={\bar m}$ and variance 
$\langle (m_r-{\bar m})(m_{r'}-{\bar m})\rangle_m=g\delta_{r,r'}$.
The following transport calculations will be based entirely on the Hamiltonian of 
Eq. (\ref{ham01}). 
In particular, the average Hamiltonian $\langle H\rangle_m$
can be diagonalized by Fourier transformation and is
$
k_1\sigma_1+k_2\sigma_2+{\bar m}\sigma_3
$ for MLG with eigenvalues $E_k=\pm\sqrt{{\bar m}^2+k^2}$.
For BGL the average Hamiltonian is
$ 
(k_1^2-k_2^2)\sigma_1+2k_1k_2\sigma_2+{\bar m}\sigma_3
$ with eigenvalues $E_k=\pm\sqrt{{\bar m}^2+k^4}$.

{\it symmetries:}
Transport properties are determined by the model properties on large scales.
The latter are controlled by the symmetry of the Hamiltonian and of the corresponding 
one-particle Green's function $G(i\epsilon)=(H+i\epsilon)^{-1}$. In the absence of
sublattice-symmetry breaking (i.e. for $m=0$), the Hamiltonian 
$H=h_1\sigma_1+h_2\sigma_2$ has a continuous chiral symmetry
\beq
H \to e^{\alpha\sigma_3} He^{\alpha\sigma_3}=H
\label{contsymmetry}
\eeq
with a continuous parameter $\alpha$, since $H$ anticommutes with $\sigma_3$.
The SP term $m\sigma_3$ breaks the continuous chiral symmetry. 
However, the behavior under transposition $h_j^T=-h_j$ for MLG and $h_j^T=h_j$ for 
BLG provides a discrete symmetry:
\beq
H\to -\sigma_j H^T\sigma_j =H \ ,
\label{discretesymm}
\eeq
where $j=1$ for MLG and $j=2$ for BLG.
This symmetry is broken for the one-particle Green's function $G(i\epsilon)$
by the $i\epsilon$ term. To see whether or not the symmetry is recovered for $\epsilon\to0$,
the difference
\beq
G(i\epsilon)+\sigma_jG^T(i\epsilon)\sigma_j=G(i\epsilon)-G(-i\epsilon)=i\pi\rho(E=0) 
\label{op}
\eeq
must be evaluated, where $\rho(E=0)\equiv\rho_0$ is the density of states at the NP.
Here the limit $\epsilon\to0$ is implicitly assumed.
Thus the order parameter for spontaneous symmetry breaking is $\rho_0$.

{\it conductivity:}
The conductivity can be calculated from the Kubo formula.
Here we focus on interband scattering between states of energy $\omega/2$ and $-\omega/2$, 
which is a major contribution to transport near the NP. The frequency-dependent 
conductivity then reads \cite{ziegler08}
\beq
\sigma_0(\omega)
=-\frac{e^2}{2h}\omega^2 \langle\langle \Phi_{-\omega/2}|r_k^2|\Phi_{\omega/2}\rangle\rangle_m \ ,
\label{cond0b}
\eeq
where $|\Phi_{E}\rangle$ is an eigenstate of $H$ in Eq. (\ref{ham01}) with energy $E$.
In other words, the conductivity is proportional to a matrix element of the position operator $r_k$
($k=1,2$) with respect to energy eigenfunctions from the lower and the upper band. 
The matrix element $\langle\Phi_{\omega/2}|r_k^2|\Phi_{-\omega/2}\rangle$ is identical with 
the two-particle Green's function
\beq
\sum_{r} r_k^2 Tr_2\left[
G_{r0}(-\omega/2-i\epsilon)G_{0r}(\omega/2+i\epsilon)\right] \ .
\label{cond2}
\eeq
This indicates that transport properties are expressed by the two-particle 
Green's function
$G(i\epsilon)G(-i\epsilon)$. Each of the two Green's functions, $G(i\epsilon)$
and $G(-i\epsilon)$, can be considered as a random variable which
are correlated due to the common random variable $m_r$. Their distribution is
defined by a joint distribution function $P[G(i\epsilon),G(-i\epsilon)]$. 
In terms of transport theory, both Green's functions must be included
on equal footing. This is possible by introducing the extended Green's function
\[
{\hat G}(i\epsilon)=\pmatrix{
G(i\epsilon) & 0 \cr
0 & G(-i\epsilon) \cr
} =\pmatrix{
H+i\epsilon & 0 \cr
0 & H-i\epsilon \cr
}^{-1} \ .
\]
In the present case one can use
the symmetry transformation of $H$ in Eq. (\ref{discretesymm}) to write 
the extended Green's function as
\[
\pmatrix{
\sigma_0 & 0 \cr
0 & i\sigma_j \cr
}\pmatrix{
H+i\epsilon & 0 \cr
0 & H^T+i\epsilon \cr
}^{-1} \pmatrix{
\sigma_0 & 0 \cr
0 & i\sigma_j \cr
} \ .
\]
This introduces an extended Hamiltonian ${\hat H}=diag(H,H^T)$ which is invariant 
under a global ``rotation'' 
\beq
{\hat H}\to e^{S}{\hat H}e^{S}={\hat H} \ , \ \ \ \ S=\pmatrix{
0 & \alpha \sigma_j \cr
\alpha'\sigma_j & 0 \cr
}
\label{symmetry2}
\eeq
with continuous parameters $\alpha,\alpha'$, since ${\hat H}$ anticommutes with $S$.
The $i\epsilon$ term of the Green's function also breaks this symmetry.
According to Eq. (\ref{op}), the symmetry is broken spontaneously  for $\epsilon\to0$
if the density of states $\rho_0$ is nonzero. Since this is a continuous symmetry, there is a 
massless mode which describes diffusion \cite{ziegler97}. Symmetry breaking should be
studied for average quantities. Therefore, the average density of states must be evaluated.

{\it spontaneous symmetry breaking:}
The average one-particle Green's function can be calculated from the average Hamiltonian 
$\langle H\rangle_m$ by employing the self-consistent Born approximation (SCBA)
\cite{suzuura02,peres06,koshino06}
\beq
\langle G(i\epsilon)\rangle_m\approx (\langle H\rangle_m - 2\Sigma)^{-1}
\equiv G_0(i\eta,m_s) \ .
\label{scba1}
\eeq
The self-energy $\Sigma$ is a $2\times2$ tensor due to the spinor structure
of the quasiparticles: $\Sigma=-(i\eta\sigma_0+m_s \sigma_3)/2$.
Scattering by the random SP 
produces an imaginary part of the self-energy $\eta$ (i.e. a one-particle scattering rate)
and a shift $m_s$ of the average SP ${\bar m}$ 
(i.e., ${\bar m}\to m'\equiv {\bar m}+m_s$). 
$\Sigma$ is determined by the self-consistent equation
\beq
\Sigma=-g\sigma_3(\langle H\rangle_m -2\Sigma)^{-1}_{rr}\sigma_3 \ .
\label{spe00}
\eeq
For simplicity, the dc limit $\omega\sim0$ is considered here.
The average density of states at the NP is proportional to the scattering rate:
$\rho_0=\eta/2g\pi$. This reflects that scattering by the random SP 
creates a nonzero density of states at the NP. It should be noticed that
the entire calculation of the one-particle scattering rate $\eta$ is based on
the average one-particle Green's function. Therefore, it is unrelated to the
continuous symmetry of Eq. (\ref{symmetry2}). On the other hand, $\eta>0$
implies spontaneous breaking of this symmetry.

Eq. (\ref{spe00}) can also be written in 
terms of two equations, one for the one-particle scattering rate $\eta$ and 
another for the shift of the SP $m_s$, as 
\beq
\eta= gI\eta, \ \ m_s=-{\bar m} gI/(1+gI) \ .
\label{scba2}
\eeq
$I$ is a function of ${\bar m}$ and $\eta$ and also depends on the Hamiltonian. 
For MLG it reads with momentum cutoff $\lambda$
\begin{equation}
I= 
\frac{1}{2\pi}\ln\left[ 1+\frac{\lambda^2}{{\eta}^2 +({\bar m}+m_s)^2}\right]
\label{int1}
\end{equation}
and for BLG
\begin{equation}
I\sim \frac{1}{4\sqrt{{\eta}^2+({\bar m}+m_s)^2}}\ \ \ \ (\lambda\sim\infty) \ .
\label{int2}
\end{equation}
A nonzero solution $\eta$ requires $gI=1$ in the first part of Eq. (\ref{scba2}), 
such that $m_s=-{\bar m}/2$ from the second part. 
Since the integrals $I$ are monotonically decreasing functions for large ${\bar m}$,
a real solution with $gI=1$ exists only for $|{\bar m}|\le m_c$. For both, MLG and BLG,
the solutions read
\beq
\eta^2=(m_c^2-{\bar m}^2)\Theta(m_c^2-{\bar m}^2)/4 \ ,
\label{scattrate}
\eeq
where the model dependence enters only through the critical average SP $m_c$:
\beq
\frac{2\lambda}{\sqrt{e^{2\pi/g}-1}}\sim 2\lambda e^{-g/\pi} \ \ (MLG),\ \ \ 
g/2\ \ (BLG) \ .
\label{gap11}
\eeq
$m_c$ is much bigger for BGL (cf. Fig. 1), a result which indicates that the effect of disorder 
is much stronger in BLG. This is also reflected by the scattering rate at ${\bar m}=0$ which 
is $\eta=m_c/2$.

{\it diffusion: 
}
The average two-particle Green's function 
\[
K^{-1}_{rr'}(i\epsilon)=-\langle Tr_2[G_{rr'}(-i\epsilon)G_{r'r}(i\epsilon)]\rangle_m
\]
can be evaluated from an effective field theory \cite{ziegler97}.  If $\eta>0$ the corresponding 
spontaneous breaking of the symmetry in Eq. (\ref{symmetry2}) creates one massless
mode, which is related to a diffusion propagator in Fourier space:
\[
\frac{1}{K_q(i\epsilon)}\sim -\frac{\eta/g}{\epsilon+D q^2}
\]
with the diffusion coefficient
\beq
D= g\frac{\eta}{2}\sum_r r_k^2 Tr_2[G_{0,r0}(-i\eta)G_{0,0r}(i\eta)] \ .
\label{diffcoeff}
\eeq
Within this approximation the matrix element of the position operator reads
\beq
\langle\langle\Phi_{\omega/2} |r_k^2|\Phi_{-\omega/2}\rangle\rangle_m
=\frac{\partial^2}{\partial q_k^2}\frac{1}{K_q(\omega/2)}\Big|_{q=0}
\sim -8\frac{\eta}{g\omega^2}D \ .
\label{matrixelement}
\eeq
Using the relation between the matrix element and the two-particle Green's function
in Eq. (\ref{cond2}), the diffusion coefficient becomes 
$D=(g\eta/2)\langle\Phi_{i\eta} |r_k^2|\Phi_{-i\eta}\rangle$.
Inserting this on the right-hand side of Eq. (\ref{matrixelement}) gives a simple relation 
between the disorder averaged matrix element of $r_k^2$ and the corresponding matrix
element without disorder:
\beq
\langle\langle\Phi_{\omega/2} |r_k^2|\Phi_{-\omega/2}\rangle\rangle_m
=-\frac{\eta^2}{(\omega/2)^2}\langle\Phi_{i\eta} |r_k^2|\Phi_{-i\eta}\rangle
\ .
\label{mespm}
\eeq
This is similar to the relation of the average one-particle Green's function in
the SCBA of Eq. (\ref{scba1}). Like in the latter case,
the averaging process leads to a change of energies $\omega/2\to i\eta$ (i.e. a replacement
of the frequency by the scattering rate). Moreover, in the relation of the two-particle Green's function
there is an extra prefactor $-\eta^2/(\omega/2)^2$. It is important for the transport properties, 
since the average matrix element diverges like $\omega^{-2}$. 
This indicates that the states $|\Phi_{\pm\omega/2}\rangle$
are delocalized for $\omega=0$ in the presence of weak SP disorder, and localization increases
as one goes away from the NP. Such a behavior was also found for bond disorder in analytic \cite{ziegler08}
and in numerical studies \cite{xiong07}.

\begin{figure}[ht]
\includegraphics[width=7cm,height=6cm]{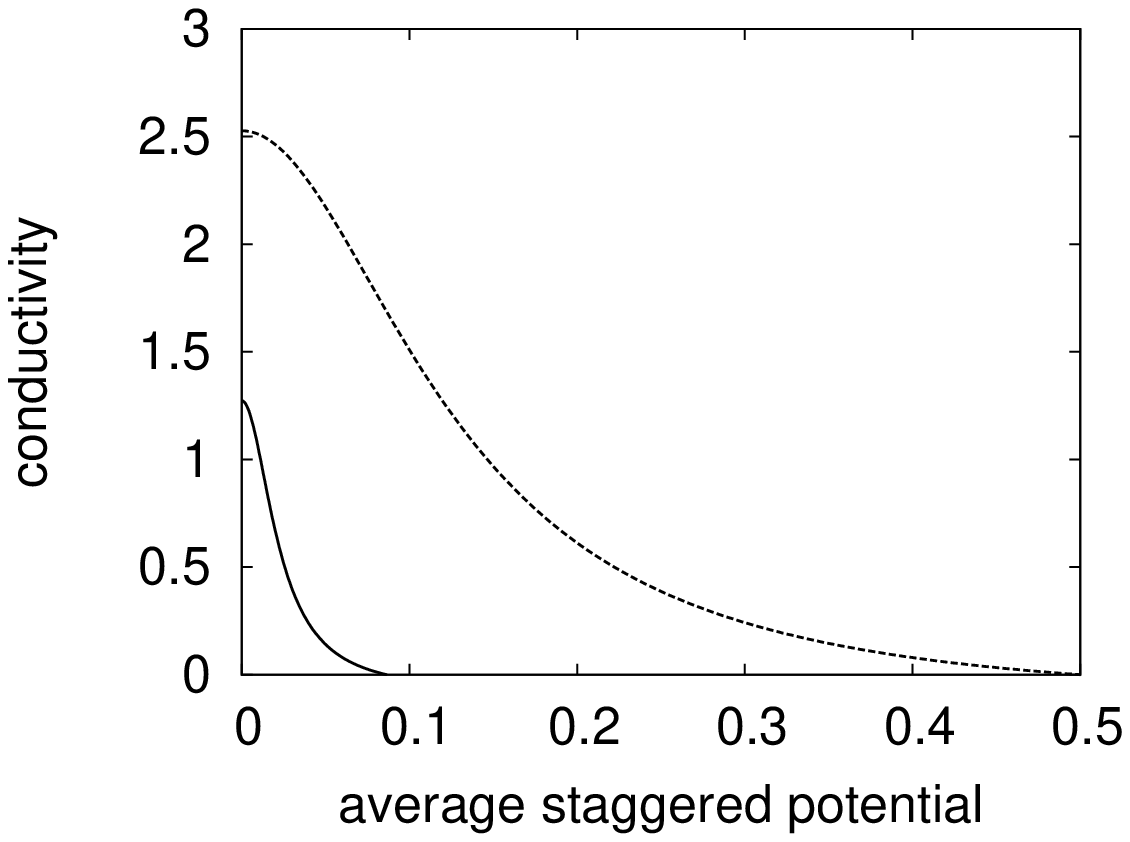}
\includegraphics[width=7cm,height=6cm]{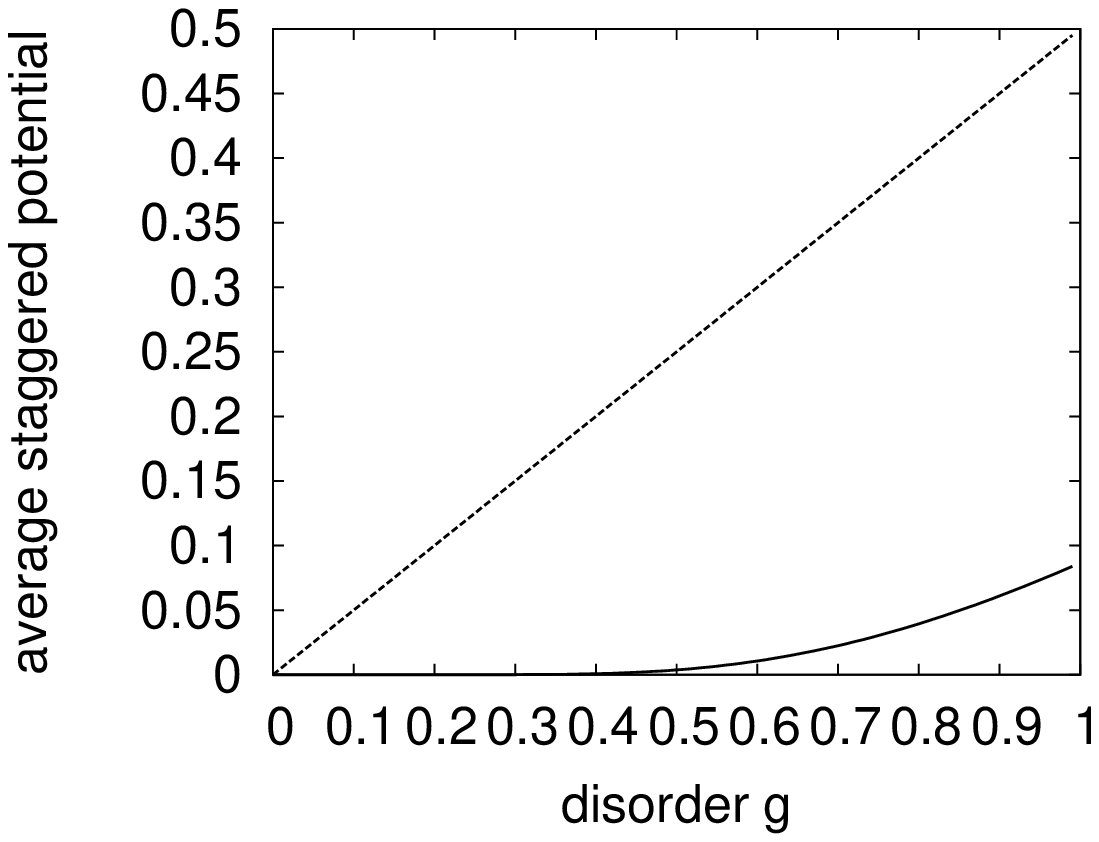}
\caption{
upper panel: dc conductivity in units of $e^2/h$ for BL graphene (upper curve) 
and ML graphene (lower curve) vs. the average staggered potential ${\bar m}$, calculated from Eq. (\ref{result})
for $g=1$ and $\lambda=1$.
lower panel: critical average staggered potential as a function of $g$ 
(variance of the staggered potential fluctuations) for BL graphene (upper curve)
and ML graphene (lower curve).}
\label{diff}
\end{figure}

After evaluating $\langle\Phi_{i\eta} |r_k^2|\Phi_{-i\eta}\rangle$,
the results for the diffusion coefficient in Eq. (\ref{diffcoeff})  and 
for the conductivity in Eqs. (\ref{cond0b}), (\ref{matrixelement})  can be summerized 
in the following expressions
\beq
D\sim 
\frac{ag}{4\pi}\frac{\eta}{\eta^2+{\bar m}^2/4}\ ,\ \ \ 
\sigma_0\sim \frac{a}{\pi}\frac{\eta^2}{\eta^2+{\bar m}^2/4}\frac{e^2}{h} \ ,
\label{result}
\eeq
where $a=1$ ($a=2$) for MLG (BLG).
First, this result indicates that the physical relevant quantity is the one-particle 
scattering rate $\eta$. 
The difference between MLG and BLG is only due to the parameter $a=1,2$ and due to the 
${\bar m}$-dependent scattering rate $\eta$. 
Second, the result reflects a diffusive behavior as long as the scattering rate $\eta$ does not
vanish. Eq. (\ref{scattrate}) gives a vanishing scattering rate for ${\bar m}=m_c$, where
the critical value $m_c$ is twice the scattering rate at ${\bar m}=0$. Moreover, the
average density of states at the NP is proportional to $\eta$. Therefore, a
global gap opens only for ${\bar m}>m_c$. 
Details of the transport properties distinguish between ${\bar m}=0$ and ${\bar m}\ne 0$.
 
${\bar m}=0$:
A fluctuating SP with $g$ not too large has no effect on the conductivity.
This can also be understood from the Einstein relation $\sigma_0\propto D \rho$,
since the density of states $\rho$ is proportional and the diffusion coefficient $D$ is
inversely proportional to the normalized scattering rate $\eta/g$. 
Such a behavior was also observed in the chiral-invariant case with random
bond disorder which is related to ripples \cite{morozov06,geim07,peres06,ziegler08}.
The scattering rate $\eta$ increases with disorder strength $g$ (cf. Fig. 1). 
Consequently, the density of states 
at the NP $\rho_0=\eta/2g\pi$ increases with $g$ for MLG, at least for small values of $g$, 
whereas it is constant for BLG. On the other hand, the diffusion coefficient decreases with 
$g/\eta$, as a result of increased scattering.

${\bar m}\ne 0$:
The conductivity decreases with ${\bar m}$ and eventually goes to zero at ${\bar m}=m_c$. 
This is due to two effects, namely the reduction of the 
density of states and the reduction of the diffusion coefficient with ${\bar m}$, caused by
a fluctuating gap. Since the product of the two quantities give the conductivity in the 
Einstein relation, the conductivity also decreases.

The only difference between MLG and BLG in our calculation is the linear (MLG) and the quadratic (BLG) 
spectrum. This has quantitative consequences for the conductivity: For BLG it is twice as big as for MLG 
at ${\bar m}=0$ and also decays on a larger scale for $0<{\bar m}\le m_c$, since the critical value is 
$m_c=g/2$ for BLG, whereas it is $m_c\sim \exp(-\pi/g)$ for MLG. 
As shown in Fig. 1, the conductivity of MLG vanishes at much lower values of ${\bar m}$.
Remarkable is the enormous difference of the scattering rate between the two systems at ${\bar m}=0$. 
As shown in Fig. 1, $\eta$ is practically zero for a large interval of $g$, whereas it increases
linearly with $g$ for BLG. This indicates that disorder has a much stronger effect in the latter.

Our result of the random SP represents a case that is different from random bond disorder 
(i.e. with chiral symmetry) and random scalar potential (breaks the chiral symmetry but not the 
sublattice symmetry). The former does not localize states at the NP, whereas the latter
has presumably always localized states, with a very large localization length though.
In a recent paper, Zhang et al. suggested a Kosterlitz-Thouless (KT) transition
for a long-range random potential  \cite{zhang08}. The KT transition is a phase transition 
that has no spontaneous symmetry breaking but a single massless mode in the ordered phase 
due to $U(1)$ phase fluctuations.
In the case of the random SP the situation is very different: There is spontaneous symmetry breaking
in the diffusive phase due to $\eta>0$. Moreover, the symmetry of the fluctuations in Eq. 
(\ref{symmetry2}) has two components rather than one.
Therefore, the transition to the insulating behavior due to random SP cannot be linked to the conventional 
KT transition.

A possible experimental realization of a random gap was recently observed by Adam et al. \cite{adam08}. 
It still remains to be seen whether or not the observed transition, which was studied by varying
the gate voltage at a fixed gap, can be related to a nonzero average SP. This would require a 
tuning of the gap fluctuations and measurement of the local density of states.

In conclusion, the one-particle scattering rate, the density of states, the diffusion coefficient, 
and the conductivity decrease with increasing average SP ${\bar m}$ and vanish at a critical point $m_c$. The
latter is exponentially small for MLG but proportional to disorder strength for BLG. Thus the effect
of disorder is much stronger in BLG.

\begin{acknowledgments}
This project was supported by a grant from the Deutsche Forschungsgemeinschaft
and by the Aspen Center for Physics.
\end{acknowledgments}

\end{document}